\documentclass[twocolumn]{aastex631}
\usepackage{graphicx}	
\usepackage{amsmath}	
\usepackage{amssymb}	
\usepackage{color}
\usepackage{enumitem}
\usepackage{hyperref}
\usepackage{verbatim}
\usepackage{mathrsfs}
\usepackage[all]{hypcap} 
\usepackage{afterpage}    
\usepackage{marginnote}
\usepackage{multirow}
\usepackage[mathlines]{lineno}
\usepackage{float}

\usepackage{siunitx}

\shortauthors{Liu et al.}
\graphicspath{{./}{figures/}}

\begin{document}

\title{Varstrometry for Off-nucleus and Dual Subkiloparsec AGN (VODKA): Three Quadruply Lensed Quasars at Cosmic Noon in HST and JWST}

\author[0000-0003-3673-7314]{Mingrui Liu}
\affiliation{\rm William H. Miller III Department of Physics and Astronomy, Johns Hopkins University, Baltimore, MD 21210, USA \\}
\author[/0000-0002-9932-1298]{Yu-Ching Chen}
\affiliation{\rm William H. Miller III Department of Physics and Astronomy, Johns Hopkins University, Baltimore, MD 21210, USA \\}
\author[0000-0001-6100-6869]{Nadia Zakamska}
\affiliation{\rm William H. Miller III Department of Physics and Astronomy, Johns Hopkins University, Baltimore, MD 21210, USA \\}
\author[0000-0003-0049-5210]{Xin Liu}
\affiliation{\rm Department of Astronomy, University of Illinois at Urbana-Champaign, Urbana, IL 61801, USA \\}
\author[0000-0003-1659-7035]{Yue Shen}
\affiliation{\rm Department of Astronomy, University of Illinois at Urbana-Champaign, Urbana, IL 61801, USA \\}
\author[0000-0001-8917-2148]{Xuheng Ding}
\affiliation{\rm School of Physics and Technology, Wuhan University, Wuhan 430072, China \\}
\author[0000-0001-7681-9213]{Arran C. Gross}
\affiliation{\rm Department of Astronomy, University of Illinois at Urbana-Champaign, Urbana, IL 61801, USA \\}
\author[0000-0003-4250-4437]{Hsiang-Chih Hwang}
\affiliation{\rm Institute for Advanced Study, Einstein Dr., Princeton, NJ 08540, USA \\}
\author[0000-0001-7572-5231]{Yuzo Ishikawa}
\affiliation{\rm Kavli Institute for Astrophysics and Space Research, Massachusetts Institute of Technology, Cambridge, MA 02139, USA \\}
\author[0000-0001-7946-557X]{Kedar Phadke}
\affiliation{\rm Department of Astronomy, University of Illinois at Urbana-Champaign, Urbana, IL 61801, USA \\}



\begin{abstract}
We present results from imaging observations of three quadruply lensed quasars by Hubble Space Telescope (HST) and James Webb Space Telescope (JWST) at redshifts $z = 2.550$, 2.975, and 1.500.
We model our targets assuming a singular isothermal ellipsoid mass profile and an elliptical Sérsic profile for the lensing galaxies, and reconstruct the geometric configuration of each system with measured Einstein radii of 0.44$''$, 0.58$''$, and 0.49$''$. While no spectroscopic measurements are available for the lenses, we constrain the redshift of each lens to $0.5 < z < 1.2$, $1.0 < z < 1.5$, and $0.4 < z < 0.9$. For all three lenses, the best-fit light models yield a typical de Vaucouleurs $n_{\rm S\acute{e}rsic} \sim 4$ profile and an effective radius $R_e$ around $\sim 1.5 - 3.5$ kpc. We accordingly classify the three lenses as early-type galaxies at an intermediate to high redshift, a common type for strong lensing galaxies. Compared to other known quadruple lenses, the lensing galaxies in this work are at the lower end of the distribution of Einstein radii and upper end of the distribution of the lens redshifts. They represent an interesting quadrant of subarcsecond-separation lenses in the population of single-galaxy strong lensing which have been largely unexplored yet and will be great targets of interest in upcoming high-resolution lensing surveys.

\end{abstract}



\section{Introduction} \label{sec:intro}
Ever since the first detection of strong gravitational lensing \citep{1979Natur.279..381W}, this relativistic phenomenon has been offering a unique perspective to various astrophysical and cosmological enigmas. For one, strong lensing serves as an effective probe for precision cosmology \citep{2010ARA&A..48...87T, 2024SSRv..220...12S}. Time delay between photons from different lensed images enables high precision measurements on the Hubble constant $H_0$, the expansion rate of the universe \citep{2010ApJ...711..201S, 2016A&ARv..24...11T}. The $H_0$ value measured in the local Universe from distance ladders such as Type Ia supernovae \citep{2016ApJ...826...56R, 2022ApJ...934L...7R} exhibits a $>$3$\sigma$ discrepancy with the value for early-stage Universe extrapolated from cosmic microwave background measurements \citep{2020A&A...641A...6P}. Whether this tension originates from systematic uncertainties or implies new physics beyond the standard cosmic acceleration ($\Lambda$) with cold dark matter ($\Lambda$CDM) paradigm, such as exotic species of particle (\citealt{2020PhRvD.101d3533K}) or early dark energy \citep{2023ARNPS..73..153K}, is still under investigation. As an independent probe for $H_0$, strong lensing offers precious insights about the nature of this ongoing Hubble tension \citep{2020MNRAS.498.1420W,2020A&A...643A.165B,2020MNRAS.494.6072S,2025A&A...704A..63T}. 

Strong lensing provides a unique laboratory for understanding galaxy structure and evolution. Lensing, often combined with stellar kinematics and photometry, probes mass structures of lensing galaxies, capable of decomposing them into stellar and dark matter constituents \citep{2014MNRAS.438.3594D, 2015ApJ...804L..21C}. It provides unique constraints on the mass distribution of galaxies at intermediate to high redshifts
\citep{2006ApJ...649..599K, 2024SSRv..220...87S}. Large lens surveys such as the Sloan Lens ACS Survey \citep{2006ApJ...638..703B} and the BOSS Emission-Line Lens Survey \citep{2012ApJ...757...82B} have found the density profile of early-type galaxies generally follows a nearly isothermal power-law, which is consistent with the $\Lambda$CDM paradigm and reveals invaluable details into galaxy evolution across cosmic time \citep{2009ApJ...703L..51K, 2024A&A...690A.325S}.

Strong lensing could be a potential tracer for dark matter in galaxies. Presence of dark matter substructures at sub-galactic scales and along the line of sight affects the geometric configuration of strong lenses, manifested as abnormal positions or flux ratios of lensed images \citep{1998MNRAS.295..587M, 2002ApJ...572...25D, 2024MNRAS.530.2960N, 2024MNRAS.535.1652K}.
Microscopic physics of dark matter could imprint observable signatures in strong lensing systems. For example, self-interacting dark matter \citep{2014PhRvL.113b1302K, 2021MNRAS.501.4610R} would modify the central density profile of lensing galaxy halos by conducting heat from the outskirts to the halo center that helps form a constant density
core \citep{2020PhRvD.101f3009N, 2023PhRvD.107j3008G}. Fuzzy dark matter \citep{2000PhRvL..85.1158H, 2014NatPh..10..496S}
could create unique granular density core at the center of lensing halos, which would be distinct from that of any other kind of dark matter particle \citep{2020PhRvL.125k1102C, 2022MNRAS.517.1867L}.

By early 2020s, a few thousands strong lenses had been discovered \citep{2019A&A...625A.119M, 2020ApJ...894...78H}.
Many detections of strong lensing systems were serendipitous in surveys not designated for lenses. One population within which strong lenses often hide is the dual active galactic nuclei (AGNs), which are proposed progenitors of supermassive black hole (SMBH) mergers \citep{1980Natur.287..307B, 2021NatAs...5..569S, 2023ApJ...943...38S}. Before two SMBHs eventually coalesce following a galaxy merger, they spend hundreds million years being separated on sub-galactic scales ($\leq$ 10 kpc; \citealt{2008ApJS..175..356H, 2022ApJ...925..162C}). If both SMBHs are activated, a pair of AGN can then be observed simultaneously (e.g. \citealt{2006ApJ...646...49R, 2010ApJ...715L..30L, 2011ApJ...735...48S}). Such objects appear as two or more closely separated point sources and are difficult to distinguish from single quasars strongly lensed by a foreground galaxy.
Varstrometry for Off-nucleus and Dual Subkiloparsec AGN (VODKA; \citealt{2019ApJ...885L...4S, 2020ApJ...888...73H}) is a dedicated search program for dual quasars, with the goal to identify dual AGNs at high redshifts ($z > 1.5$) with subarcsecond separation from large all-sky surveys such as Gaia and Sloan Digital Sky Survey (SDSS). 
To date, VODKA has identified 150 dual AGN candidates, with at least 15 of them suspected or already confirmed to be strong lenses \citep{2022ApJ...925..162C, Gross_2025}.

\begin{figure}
\label{fig.hst_b}
    \centering
    \includegraphics[width=1.0\linewidth]{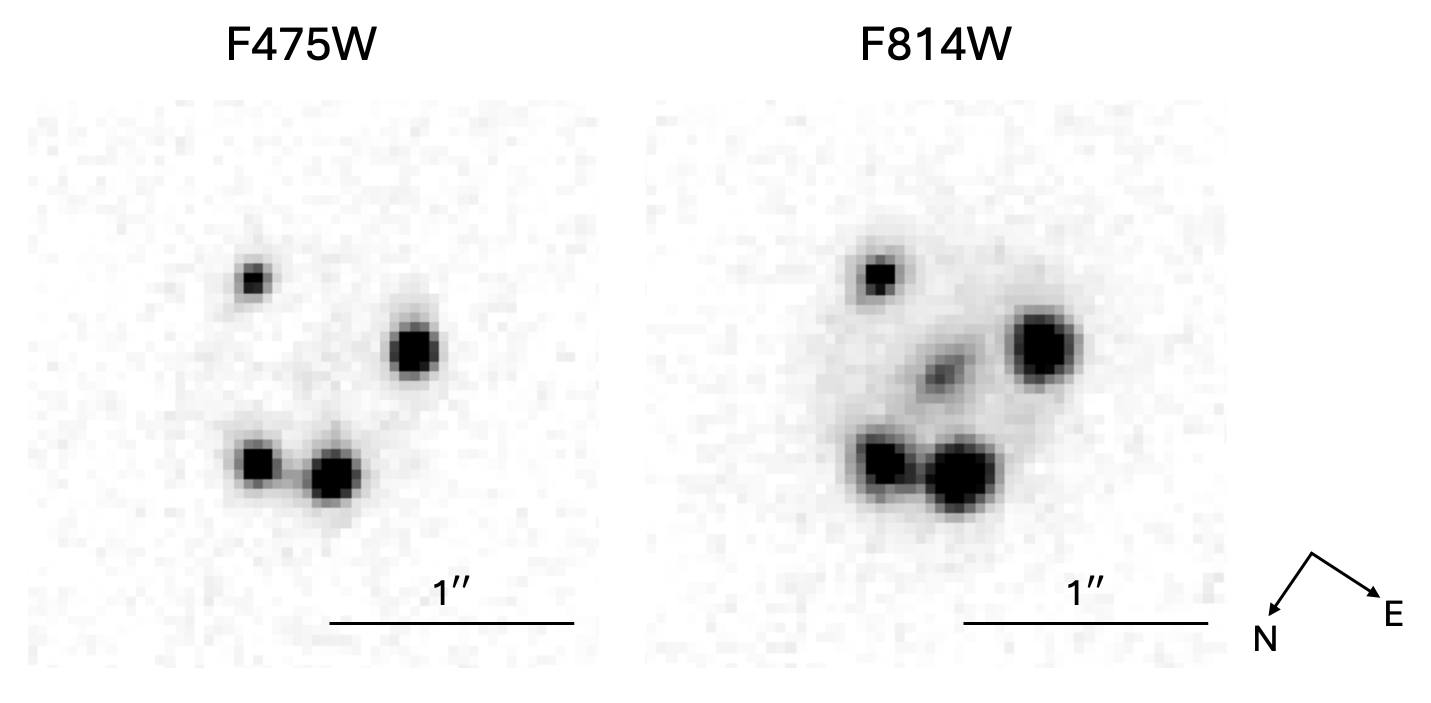}
    \caption{Images of J2218-3322 in the HST WFC3 F475W and F814W band. The arrows show orientation on the sky in this and all subsequent images.}
\end{figure}

\begin{figure*}
\label{fig.jwst1_b}
    \centering
    \includegraphics[width=1.0\linewidth]{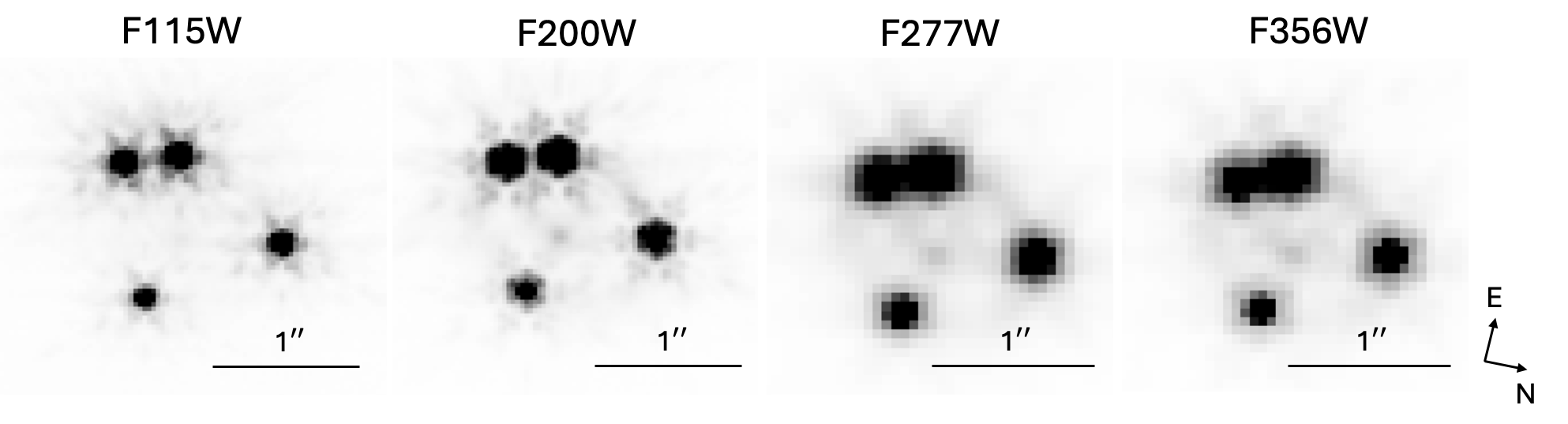}
    \caption{Images of J0803+3908 in the JWST NIRCam F115W, F200W, F277W, and F356W band.}
\end{figure*}

\begin{figure*}
\label{fig.jwst2_b}
    \centering
    \includegraphics[width=1.0\linewidth]{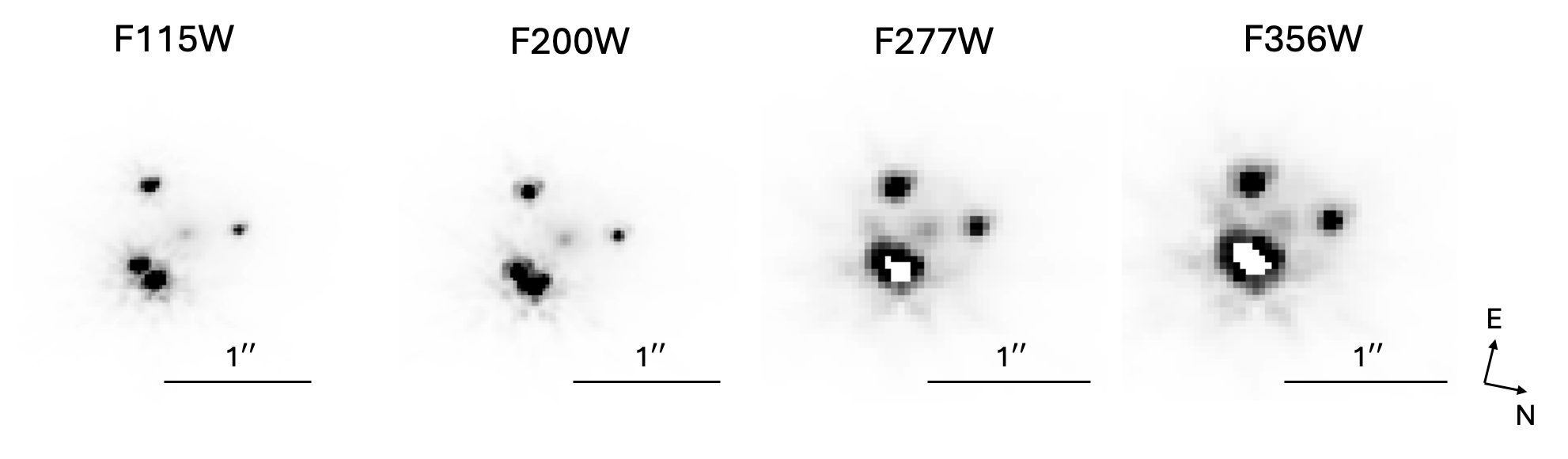}
    \caption{Images of J0813+2545 in the JWST NIRCam F115W, F200W, F277W, and F356W band. Voids in the two long filters are saturated pixels blanked out during data reduction.}
\end{figure*}

In this work, we report imaging analysis of three quadruply lensed quasars: J221849.85-332244.2, MQJ080357.75+390823.0, and SDSSJ081331.27+254503.1 (hereafter J2218-3322, J0803+3908, and J0813+2545, respectively).
We present detailed modeling and analysis of these three quadruply lensed systems in the following order:
in Section~\ref{sec:obs}, we introduce the selection and data reduction criteria for observations used in this work. In Section~\ref{sec:mod}, we outline the details of our numerical modeling conducted. We then summarize the principal scientific findings in Section~\ref{sec:analy} and discuss their implications in Section~\ref{sec:dis}. A final conclusion is drawn in Section~\ref{sec:con}. For this work, we adopt a flat $\Lambda$CDM cosmology with $\Omega_{\Lambda} = 0.7$, $\Omega_{M} = 0.3$, and $H_0 = 70$ km s$\rm ^{-1}$ Mpc$\rm ^{-1}$. All magnitudes are expressed in the AB system.

\section{Observations \& Data Reduction} \label{sec:obs}
\subsection{Target Selection}
One of our targets, J2218-3322, was observed with the Wide Field Camera 3 (WFC3) on HST. The other two sources, J0803+3908 and J0813+2545, were observed with the Near Infrared Camera (NIRCam) on JWST. 

J2218-3322 was selected to be observed by HST initially as a dual quasar candidate in the VODKA program. It was identified as a photometric quasar in both Wide-field Infrared Survey Explorer (WISE; \citealt{2012MNRAS.426.3271M, 2015ApJS..221...12S}) and Panoramic Survey Telescope and Rapid Response System Telescope \#1 (Pan-STARRS1; \citealt{2016arXiv161205560C}) data. However, in Gaia this target matches with two sources within 3$"$ of each other, with a Gaia G-band magnitude of 20.66 mag and 20.17 mag, as well as a Gaia \texttt{astrometric\_excess\_noise} property of 7.320 mas and 5.016 mas. As a result, \cite{2022ApJ...925..162C} identify it as a candidate dual or lensed quasar and use HST imaging to demonstrate that the morphology is consistent with a quadruple lens.
With optical spectra from Gemini Multi-Object Spectrographs and Space Telescope Imaging Spectrograph (STIS) on HST, \cite{2025ApJ...988..126C} determine a spectroscopic redshift of the lensed quasar (source) to be $z = 2.550$. The lensing galaxy of this system has not been analyzed in any previous work, and no spectroscopic or photometric redshift of it is available at present.

J0803+3908 and J0813+2545 are known quadruply lensed quasars. They are targets in the JWST Cycle 2 program GO-4204 (PI: Chen) to investigate small-separation dual quasars and lensed quasars. J0803+3908 is a confirmed lensed quasar in Gaia DR2 found by the dedicated survey for gravitationally lensed quasars in the Gaia catalog first conducted by \cite{2017MNRAS.472.5023L}. \cite{2018ApJ...863..144S} report a spectroscopic redshift of $z = 2.975$ for the source quasar. J0813+2545 is a quad lens found serendipitously by \cite{2002A&A...382L..26R} (who refer to this target as HS 0810+2554) when target acquisition images on HST/STIS captured the signature lensing configuration of this system. The modern quality of NIRCam imaging data compensates for the limited resolution and short exposure time (12 seconds) of the original images, thus better serving our scientific goal. The Hamburg All Sky Bright QSO Surveys \citep{1995A&AS..111..195H, 1997Msngr..88...14R} report a spectroscopic redshift of the source quasar at $z = 1.500$ \citep{1999A&AS..134..483H, 2002A&A...382L..26R}. Recently this system has been observed in the radio C and Ku bands by the Very Large Array as a part of the VLA/24B-361 program (PI: Gross). Corresponding analysis (coming up in Baltaci et al. 2026, in prep.) indicates that in both bands the lensing geometry is seen as expected, but in the C-band there is blending between the two closest quasar images. No radio emission from the central lensing galaxy is detected. Redshift of the lens galaxy in neither target is known.

\begin{figure*}
\label{fig.hst_f}
    \centering
    \includegraphics[width=1.0\linewidth]{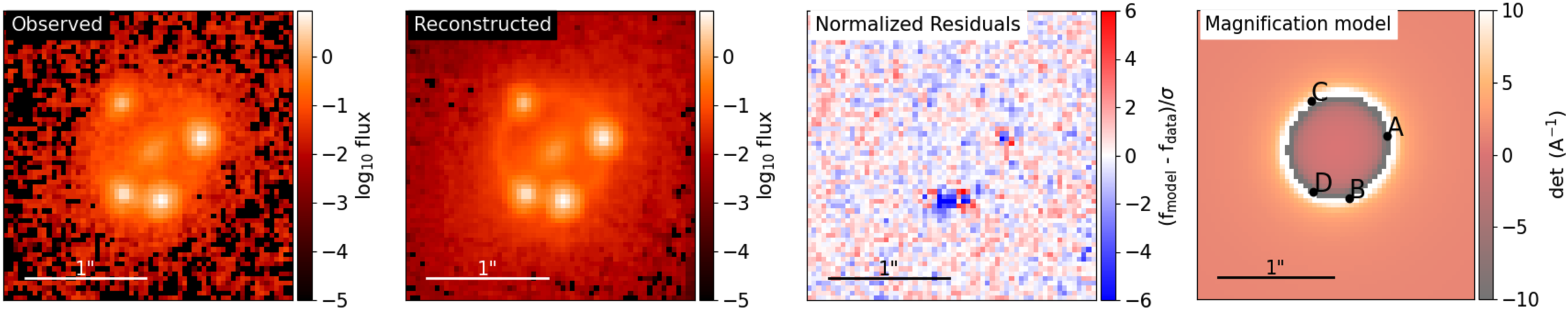}
    \caption{Best-fit models for J2218-3322 with \texttt{Lenstronomy} in the F814W band. Each panel from left to right presents the observed image, the reconstructed image, the normalized residual map, and the magnification map.}
\end{figure*}

\begin{table*}
\centering
\label{tab:para_combined}
\begin{tabular}{lccc}
\hline
Parameter & J2218-3322 & J0803+3908 & J0813+2545 \\
\hline
$\theta_E$ ($''$) & $0.437^{+0.001}_{-0.001}$ & $0.581^{+0.001}_{-0.001}$ & $0.489^{+0.002}_{-0.002}$ \\
$R_e$ ($''$) & $0.30^{+0.02}_{-0.02}$ & $0.35^{+0.07}_{-0.03}$ & $0.45^{+0.01}_{-0.01}$ \\
$n_{\rm S\acute{e}rsic}$ & $4.24^{+0.30}_{-0.18}$ & $4.10^{+0.08}_{-0.12}$ & $3.37^{+0.18}_{-0.09}$ \\
$q_L$ & $0.61^{+0.03}_{-0.03}$ & $0.77^{+0.02}_{-0.02}$ & $0.91^{+0.03}_{-0.03}$ \\
$\phi_L$ ($^\circ$) & $42.61^{+1.21}_{-1.21}$ & $-22.54^{+2.26}_{-2.26}$ & $13.28^{+3.22}_{-3.22}$ \\
$q_M$ & $0.94^{+0.04}_{-0.05}$ & $0.74^{+0.04}_{-0.04}$ & $0.91^{+0.06}_{-0.08}$ \\
$\phi_M$ ($^\circ$) & $28.46^{+0.47}_{-0.73}$ & $-24.12^{+3.35}_{-3.42}$ & $10.22^{+0.78}_{-1.94}$ \\
Lens $\Delta x$ ($''$) & $0.05^{+0.02}_{-0.02}$ & $-0.030^{+0.002}_{-0.002}$ & $0.01^{+0.01}_{-0.05}$ \\
Lens $\Delta y$ ($''$) & $0.06^{+0.01}_{-0.01}$ & $0.016^{+0.002}_{-0.001}$ & $0.01^{+0.01}_{-0.04}$ \\
Image A $\Delta x$ ($''$) & $0.440^{+0.002}_{-0.002}$ & $-0.252^{+0.002}_{-0.002}$ & $-0.39^{+0.02}_{-0.05}$ \\
Image A $\Delta y$ ($''$) & $0.16^{+0.01}_{-0.01}$ & $-0.374^{+0.001}_{-0.003}$ & $-0.28^{+0.04}_{-0.01}$ \\
Image B $\Delta x$ ($''$) & $0.110^{+0.002}_{-0.002}$ & $0.67^{+0.10}_{-0.10}$ & $0.47^{+0.05}_{-0.05}$ \\
Image B $\Delta y$ ($''$) & $-0.360^{+0.001}_{-0.001}$ & $-0.004^{+0.001}_{-0.001}$ & $0.03^{+0.01}_{-0.04}$ \\
Image C $\Delta x$ ($''$) & $-0.21^{+0.01}_{-0.01}$ & $-0.394^{+0.001}_{-0.001}$ & $-0.30^{+0.03}_{-0.05}$ \\
Image C $\Delta y$ ($''$) & $0.34^{+0.10}_{-0.10}$ & $0.543^{+0.002}_{-0.001}$ & $0.41^{+0.03}_{-0.04}$ \\
Image D $\Delta x$ ($''$) & $-0.20^{+0.01}_{-0.01}$ & $-0.024^{+0.001}_{-0.001}$ & $-0.26^{+0.01}_{-0.04}$ \\
Image D $\Delta y$ ($''$) & $-0.309^{+0.001}_{-0.001}$ & $0.587^{+0.003}_{-0.002}$ & $-0.41^{+0.04}_{-0.04}$ \\
\hline
\end{tabular}
\caption{Best-fit values of parameters with uncertainties for the lens mass and light models of each quadruply lensed quasar system from \texttt{Lenstronomy}. From top to bottom: Einstein radius $\theta_E$, effective radius $R_{e}$, S\'{e}rsic index $n_{\rm S\acute{e}rsic}$, light model minor-to-major axis ratio $q_L$ and position angle $\phi_L$, mass model axis ratio $q_M$ and position angle $\phi_M$, centroid positions of lens and each lensed image $\Delta x$ and $\Delta y$ (defined as displacements from image center). Uncertainties include both the fitting uncertainties (the 16th and 84th percentiles of the MCMC posterior distributions) and the run-to-run variability noted in Sec.~\ref{subsec:rob}. \texttt{Lenstronomy} is capable of achieving sub-two-decimal precision for its fits, see e.g. \cite{2019MNRAS.483.5649S, 2023ApJ...955L..16L, 2023MNRAS.518.1260S, 2025A&A...699A.259A}.}
\end{table*}

\subsection{HST Target} 
\cite{2022ApJ...925..162C} present HST/WFC3 imaging observations of J2218-3322, using the F475W (pivot wavelength $\lambda_p$ = 4772 $\rm \AA$; effective width = 1344 $\rm \AA$) and F814W ($\lambda_p$ = 8024 $\rm\AA$; effective width = 1536 $\rm \AA$) filters in the UV/Visible channel (UVIS). These data were obtained as part of  the HST Cycle 27 program SNAP-15900 (PI: Hwang), carried out on August 21$\rm ^{st}$, 2020 Universal Time (UT), with an exposure time of 360 seconds for the F475W band and 400 seconds for the F814W band. Each filter has a field of view of 20.5$^{''}$ $\times$ 20.5$^{''}$, resulting from a 512$\times$512 pixel array where each pixel is 0.04$^{''}$. The data were processed with the standard HST data reduction pipelines.  Images were first corrected for distortion and pixel effects. Next the dithered frames are cleaned from cosmic rays and hot pixels then combined, and finally calibrated both photometrically and astrometrically with \texttt{calwf3} and \texttt{MultiDrizzle}.

\subsection{JWST Targets}
JWST/NIRCam observations of J0803+3908 and J0813+2545 sources were performed as parts of the JWST Cycle 2 program GO-4204 (PI: Chen). J0803+3908 was observed with two short filters F115W ($\lambda_p$ = 1.154 $\mu$m; bandwidth $\Delta\lambda$ = 0.225 $\mu$m) and F200W ($\lambda_p$ = 1.988 $\mu$m; $\Delta\lambda$ = 0.463 $\mu$m), as well as two long filters F277W ($\lambda_p$ = 2.776 $\mu$m; $\Delta\lambda$ = 0.673 $\mu$m) and F356W ($\lambda_p$ = 3.566 $\mu$m; $\Delta\lambda$ = 0.786 $\mu$m) on April 17$\rm ^{th}$, 2024 UT. J0813+2545 was observed with the same four bands on March 29$\rm ^{th}$, 2024 UT. The nominal pixel scale is 0.031$''$/pixel for the short filters and 0.06$''$/pixel for the long filters, with the point-spread function (PSF) FWHM of 0.07$''$ at 2 $\mu$m for the short and 0.13$''$ at 4 $\mu$m for the long filters. All bands share the same field of view of 2.2$^{'}$ $\times$ 2.2$^{'}$. The data were reduced with the JWST Calibration Pipeline package in version 1.18.0 and {CRDS Context file \texttt{jwst\_1364.pmap}. 
\begin{figure*}
\label{fig.jwst1_f}
    \centering
    \includegraphics[width=1.0\linewidth]{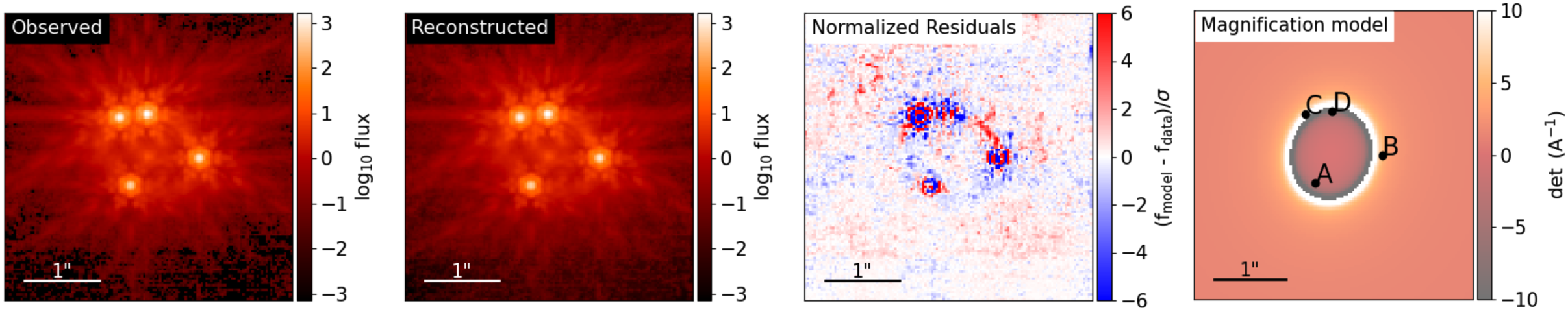}
    \caption{Best-fit models for J0803+3908 with \texttt{Lenstronomy} in the F200W band. Panels are arranged in the same convention as in Fig.~\ref{fig.hst_f}.}
\end{figure*}
We follow most of the standard procedures except for two customizations: i) the \texttt{clean\_flicker\_noise} step was turned on in Stage 1 in order to correct for the 1/f noise that produces the striping patterns in the original images; ii) the \texttt{suppress\_one\_group} property in the \texttt{ramp\_fit} step in Stage 1 was turned off to recover bright pixels at the center of certain quasar images that were saturated after integration in Frame 1 and masked out by the default pipeline settings. Turning off the flag preserves these pixels in following reduction stages and the final data product. This, however, cannot retrieve pixels saturated before Frame 1, which are the voids seen in Fig.~\ref{fig.jwst2_b}.

\section{Modeling} \label{sec:mod}
For fitting the quadruply lensed quasars, we utilize \texttt{Lenstronomy} \citep{2018PDU....22..189B}, an open-source python-based software for numerically modeling gravitational lensing systems. It offers a high level of flexibility in choices of models for the mass distribution of the lens, as well as light profiles for the lens and the source. The best-fit parameters in the models are acquired with numerical techniques such as the particle swarm optimization or Markov chain Monte Carlo algorithm. 

\subsection{Lens Mass Profile}
The mass distribution of a galaxy can often be modeled by an elliptical mass density distribution \citep{2004ApJ...611..739T, 2014MNRAS.438.3594D}.
One of such models is the singular isothermal ellipsoid (SIE; \citealt{1994A&A...284..285K}), which has been verified on a wide selection of strong lensing galaxies \citep{2008ApJ...682..964B, 2009ApJ...703L..51K, 2010ApJ...724..511A}.
This profile represents a 3D mass density profile that radially decays as a power law on ellipsoid of rotation with semi-major axes $r$: $\rho \propto r^{-\gamma}$, with a power-law slope $\gamma = 2$ \citep{2010ApJ...724..511A}. We adopt this in \texttt{Lenstronomy} to model the total mass (both baryons and dark matter) of our lens samples. In lensing-related literature, this model is often reduced to the surface mass density or ``convergence": 
\begin{equation}
\begin{aligned}
&\kappa(\theta_1, \theta_2) = \frac{3 - \gamma}{2} \left( \frac{\theta_E}{\sqrt{q\theta_1^2 + \theta_2^2/q}}\right)^{\gamma - 1} \label{eq:convergence}
\end{aligned}
\end{equation}
where $\theta_1$, $\theta_2$ are coordinates aligned with the major and minor axes of the projected ellipse rotated from sky-plane coordinates with a position angle $\phi$ of the rotated coordinate system,
$q$ is the minor-to-major axis ratio, $\theta_E$ is the Einstein radius of the lens, and power-law index $\gamma$ is fixed at 2 in our modeling \citep{2015A&A...580A..79T, 2023MNRAS.518.1260S}. This quantity is also a dimensionless ratio $\kappa = \sum/\sum_{\rm crit}$, where $\sum$ is the lens' projected mass density on the lens plane, and $\sum_{\rm crit} = \frac{c^2}{4\pi G}\frac{D_S}{D_L D_{LS}}$ is the critical surface mass density, where $D_L$, $D_S$, $D_{LS}$ are the angular diameter distance to the lensing galaxy, the source quasar, and that between the two, respectively. Multiple lensed images of the source can only be observed when the lens' $\sum$ exceeds the critical value \citep{2024SSRv..220...12S}.

In addition, other perturbers, such as galaxies or clusters either nearby or along the line of sight \citep{1988MNRAS.235.1073K}, large-scale structures \citep{1996ApJ...468...17B}, or even complicated structures within the lensing galaxy \citep{2024MNRAS.531.3684E} might disturb the lensing potential and consequently contribute to the inferred mass profile of the lens or introduce shear. We have an additional shear term which had little effects on the final goodness-of-fit and best-fit parameter values. We thus conclude that none of the quads are subject to significant external shears and report best-fit models without shears.

\begin{figure*}
\label{fig.jwst2_f}
    \centering
    \includegraphics[width=1.0\linewidth]{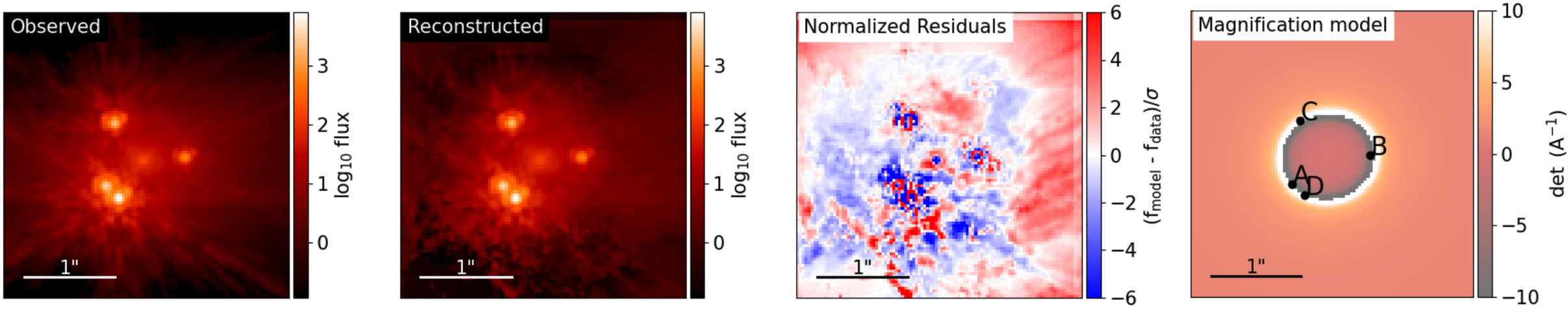}
    \caption{Best-fit models for J0813+2545 with \texttt{Lenstronomy} in the F200W band. Panels are arranged in the same convention as in Fig.~\ref{fig.hst_f}.}
\end{figure*}

\texttt{Lenstronomy} computes the magnification model by solving the lens equation with the chosen mass model and positions of the lensed images. It is not constrained by the actual flux ratios observed. When modeling the image fluxes, \texttt{Lenstronomy} can either fix them to the values predicted by this magnification model, or treat them as independent free parameters\footnote{See documentation for the lenstronomy.PointSource module: lenstronomy.readthedocs.io/en/latest/lenstronomy.PointSource.html}. Here, keeping in mind the possibility of flux ratio anomalies (as discussed in Sec.~\ref{sec:4.1.1}), we adopt the latter approach. This is enabled by setting \texttt{fixed\_magnification\_list} to \texttt{False} when specifying the lens model parameters fed into the \texttt{FittingSequence} class.

\subsection{Lens \& Source Light Profile}
For the light (i.e. surface brightness) profile of the lens and the source, we adopt \cite{1968adga.book.....S} models. Their 1D projection along the radial direction is characterized as
\begin{equation}
I(R) = I_{\rm e} \exp \left( -b_n \left[ \left(\frac{R}{R_{\rm e}}\right)^{\frac{1}{n}}-1\right]\right)
\end{equation}
with $b_{n}\approx 1.999n-0.327$, where $R_{\rm e}$ is the half-light/effective radius of the lens, $n$ is the Sérsic index, and $I_{\rm e}$ is the surface brightness at the effective radius i.e. $I_{\rm e} = I(R_{\rm e})$  \citep{2023MNRAS.518.1260S}. 

For the lens, to account for its possible ellipticity, the radius $R = \sqrt{q \theta^2_1 + \theta^2_2/q}$ features coordinates $\theta_1$ and $\theta_2$ as well as axis ratio $q$ all defined in the same conventions as in $\kappa$ above. The position angle of the ellipse is another fitted parameter in the model. 
We do not enforce these position and ellipticity terms to be identical in the mass and light profile, as we fit the two profiles separately with no assumption on whether these two profiles for our lenses trace each other in morphology. 

When axis ratio $q$ is fixed at 1, the elliptical profile reduces to a circular Sérsic model. We model the light profile of the source quasar's host galaxy with this spherical profile, and the light component of the quasar with a delta function. The Sérsic index is bounded in between 1 and 6 to avoid non-physical results.

\subsection{PSF Model}
To obtain a realistic fit of the lenses, \texttt{Lenstronomy} requires a model of the point spread function (PSF) for the instrument that produces the input images, which are convolved with the model light profiles to reproduce the observed data. For J2218-3322, due to the undersampled nature of WFC3 images, a precise analytical PSF of the detector is difficult to acquire. Rather, it is common to extract an empirical PSF model from a different object than the target in the same image, usually a nearby field star. No star is in the field of J2218-3322 that has sufficient brightness necessary for our science goal. Instead, we employ an empirical PSF extracted from an object observed by the same WFC3 filter with a similar exposure time in an archival data hosted on the Mikulski Archive for Space Telescopes (MAST; file ID: 1329728, originally from HST program GO-15986, PI: Mutchler). For J0803+3908 and J0813+2545, both fields have very few bright stars with only the brightest one showing satisfactory quality, thus we extract the brightest star in each of the fields and set them as the sample PSF model. This ensures the templates are reduced with the same procedure as that for the data.

\texttt{Lenstronomy} allows iteratively refining the PSF model during the fits \citep{2018PDU....22..189B,2025A&A...703A.118W}. We adopt this functionality in our fits for the JWST targets and briefly summarize the recipe here: first, \texttt{Lenstronomy} takes an initial PSF model and performs a preliminary fit. Based on combined residuals (via median stacking) of point sources from the preliminary fit, it updates the PSF kernel. This process is repeated for 20 iterations with only 20\% of the estimated correction applied at each step to avoid over-correction. Convergence is monitored via the log-likelihood during the updates, and iteration halts early if the reconstruction worsens. We restrict the PSF error estimation to a radius of 0.08$''$ around each point source to avoid double-counting errors from overlapping PSF wings. Upon convergence, a second fit is then performed using the updated PSF kernel, yielding the final model parameters.

\begin{figure*}
\label{fig.hst_ml}
    \centering
    \includegraphics[width=1.0\linewidth]{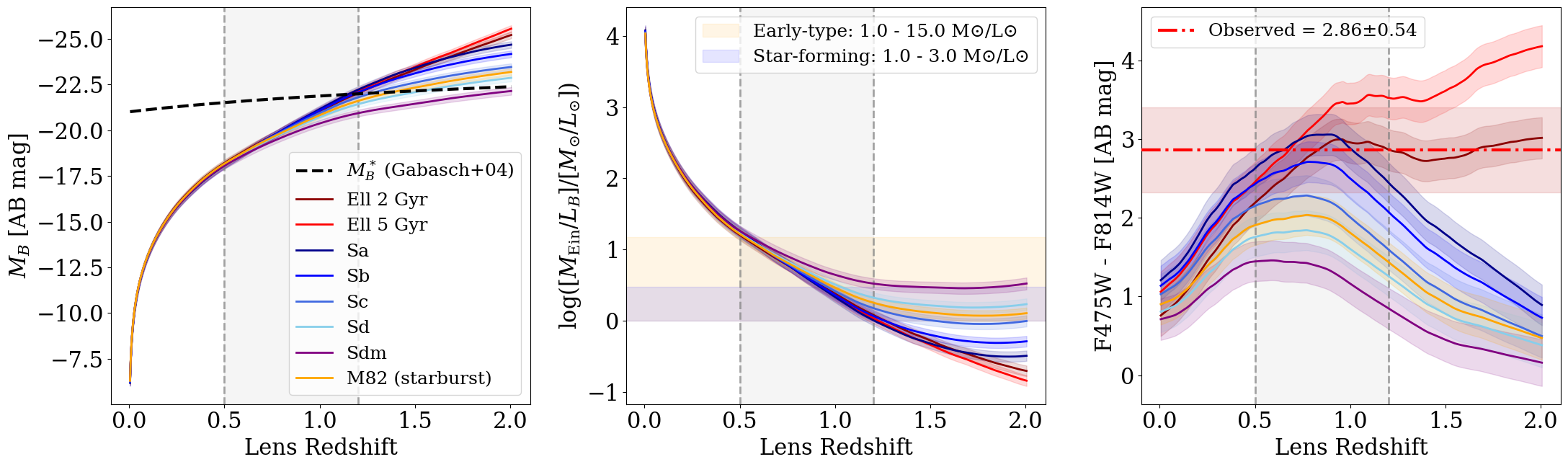}
    \caption{Photometry analysis of J2218-3322's lens. Left to right: rest-frame B-band luminosity (in AB mag), B-band mass-to-light ratio, and WFC3 [F475W - F814W] color. Black dotted line lays out B-band characteristic luminosity of early-type galaxies at intermediate redshifts reproduced from \cite{2004A&A...421...41G}. Shaded regions near each curve indicates the relative size of associated uncertainty. Vertical dotted gray lines mark the range of most probable redshift for this lens between $0.5 < z < 1.2$.}
\end{figure*}

\subsection{Noise Properties}
\texttt{Lenstronomy} searches across the parameter space for the best-fit parameters that yield the maximum likelihood, and evaluates the goodness-of-fit using the reduced chi-square parameter $\chi^2_{\nu}$, where $\chi^2_{\nu} \sim 1$ indicates a good fit. For that, it requires the noise/uncertainty in the inputted image data as an initial parameter. For HST data, we identify two dominant sources of uncertainty: the background, and the Poisson noise.
We calculate the background rms noise $\sigma_{\rm rms}$ by sampling multiple background regions (blank areas without any bright objects) across the image field and extracting the median of their root mean square (rms) values. The Poisson noise is the shot noise of the detector, which we evaluate as $\sigma_{\rm Poi} = \sqrt{N}$ with \textit{N} being the number of electrons detected during an exposure. The total uncertainty is then the quadratic sum of both noises: $\sigma_{\rm tot} = \sqrt{\sigma_{\rm rms}^2 + \sigma_{\rm Poi}^2}$.

By default, NIRCam data cubes include a data array (the \texttt{ERR} extension in the FITS file) that contains the uncertainty for each pixel value of the science product. It is calculated as the standard deviation combining variance for the Poisson noise $\sigma_{\rm Poi}^2$, variance for the detector read out noise $\sigma_{\rm r}^2$, and variance for the flat-field noise $\sigma_{\rm ff}^2$ i.e. $\sigma_{\rm ERR} = \sqrt{\sigma_{\rm Poi}^2 + \sigma_{\rm r}^2 + \sigma_{\rm ff}^2}$, to which we add the median background rms value $\sigma_{\rm rms}$ calculated by \texttt{photutils} during background subtraction in quadrature to obtain the final total uncertainty $\sigma_{\rm tot} = \sqrt{\sigma_{\rm rms}^2 + \sigma_{\rm ERR}^2}$. When the iterative PSF reconstruction is enabled, \texttt{Lenstronomy} automatically computes $\sigma_{\rm PSF}^2$ — a 2D map of the pixel-level variance in the PSF estimates, derived from the scatter among the individual PSF reconstructions at each quasar image position. This variance map enters the total noise model around the position of each point source as an additional term $\sigma_{\rm PSF}^2 \times A_{\rm ps}^2$, where $A_{\rm ps}$ is the amplitude (total flux) of each point source image\footnote{See documentation for the lenstronomy.Data.psf module: lenstronomy.readthedocs.io/en/latest/lenstronomy.Data.html}. This effectively accounts for the uncertainty induced by a PSF model.

\subsection{Model Robustness}\label{subsec:rob}
\begin{figure*}
\label{fig.jwst1_ml}
    \centering
    \includegraphics[width=1.0\linewidth]{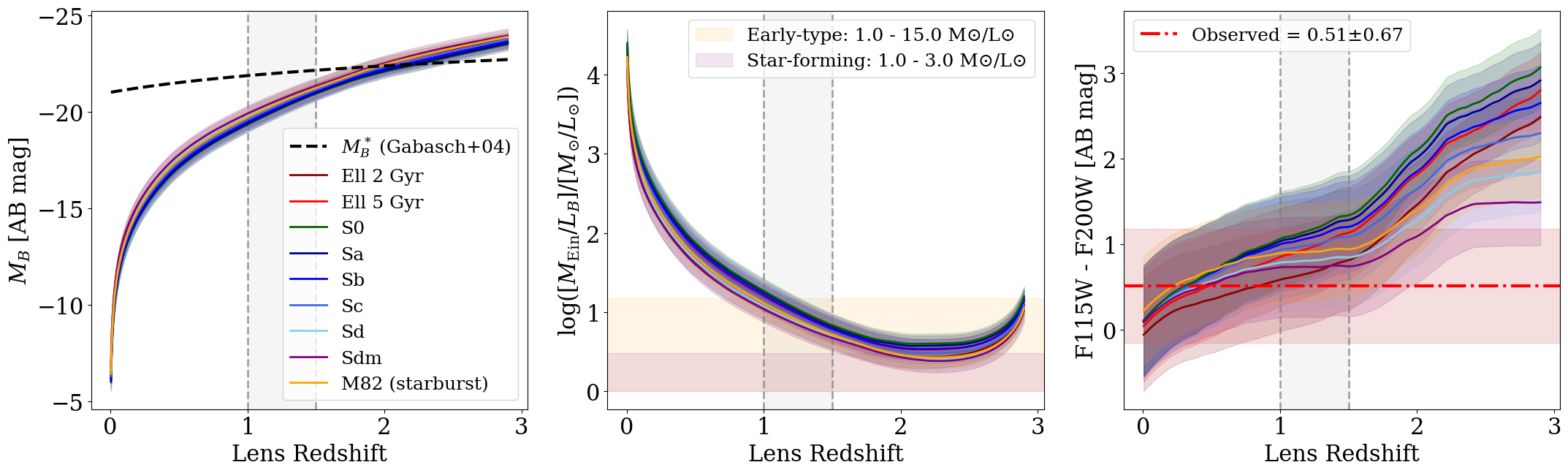}
    \caption{Photometry analysis of J0803+3908's lens. Panels are arranged in the same convention as in Fig.~\ref{fig.hst_ml} while third panel shows the NIRCam [F115W - F200W] color. Range of most probable redshift for this lens is between $1.0 < z < 1.5$.}
\end{figure*}

In addition to the SIE mass model introduced above that describes the total mass of a galaxy, we also experiment with a ``composite" model, which characterizes the baryon and dark matter in a lens with two distinct profiles. For the baryonic/luminous mass i.e. the stellar halo, we adopt a \cite{1990ApJ...356..359H} model. For the dark matter halo, we choose a Navarro-Frenk-White (NFW; \citealt{1997ApJ...490..493N}) profile. Both models in their original forms require spherical symmetry and fall short of accounting for ellipticity in realistic halos. Thus, we utilize a modified version for the two proposed by \cite{2021PASP..133g4504O}, which uses cored steep ellipsoids (CSE; \citealt{1998ApJ...495..157K, 2001astro.ph..2341K}) to approximate an elliptical Hernquist and NFW halo. Such a composite model, when converged properly, yields very similar results to that from the SIE model, including the Einstein radius, ellipticity, and centroid position of a lens. We do not include results from this alternative model because with the composite model, two consecutive trials of fitting the same target with the same set of input parameters sometimes produce significantly different outcomes, especially parameters of the stellar halo profile. 

Certain parameters are more robust than others. We observe that for the same lens being modeled, the Einstein radius is usually the most stable and rarely changes over by $\sim 0.001''$ across trials, followed by centroid positions that could change by up to $\sim 0.1''$ each trial, with ellipticity being the least robust parameter that would occasionally vary by more than $10\%$ in different trials. Fits using the SIE model demonstrate a similar hierarchy of parameter robustness: Einstein radius $>$ centroid position $>$ ellipticity. The magnitude of this run-to-run variability is included in the uncertainties reported in Table~\ref{tab:para_combined}. We warn that exact ellipticity (and possibly orientation) of the mass model might be uncertain.

\section{Analysis} \label{sec:analy}
\subsection{Imaging \& Modeling Results}
In this section we present imaging analysis and \texttt{Lenstronomy} modeling results of the three quads.

\subsubsection{J2218-3322}\label{sec:4.1.1}
Fig.~\ref{fig.hst_b} displays the WFC3 images of J2218-3322 in the F475W and F814W bands. This system consists of four point-like sources surrounding a central extended object, a classic geometry for a background quasar strongly lensed into quadruple images by a foreground galaxy. The central lensing galaxy appears as a compact object elongated along the northeast-southwest direction. This morphology is more likely being indicative of its early-type nature (discussed in Sec.~\ref{sec:dis}) than being due to lensing. It is more evident in the F814W band while barely visible in the F475W band. Enclosing the central object is a very faint Einstein ring, which again is brighter in the redder filter, and the four lensed quasar images are positioned on the Einstein ring. The brightest image is at the lower right of the field, while the faintest is at the upper left. 

Fig.~\ref{fig.hst_f} presents the \texttt{Lenstronomy} best-fit results for the system in the F814W band. We consider this band to be more informative for \texttt{Lenstronomy} as the lens has a larger flux in it and perform fitting with it. The fit produces a good reconstruction of the system that is morphologically very similar to the original image. The normalized residual map illustrates a pattern that appears as a random distribution spatially. At most pixels, the fitted flux value deviates from the actual value by $\leq 3\sigma$. This reinforces the reliability of the fit, along with the best-fit $\chi^2_{\nu} = 1.21$. The fitted magnification model suggests the point source B should be the brightest and A the dimmest, which disagrees with the data and reconstruction. Such a mismatch of magnification, so-called flux ratio anomaly, could be attributed to mass fluctuations caused by structures at sub-galactic scales that distort the lensing potential, for example, dark matter halos, for which a smooth mass profile like the SIE model cannot fully account. This might imply the existence of prominent dark matter substructures within the halo of the lens \citep{2024MNRAS.530.2960N,2024SSRv..220...58V}. Microlensing by individual stars and dust extinction could contribute to such anomalies as well \citep{2002ApJ...580..685S, 2006ApJ...639....1K}. This is a long known phenomenon in numerical modeling of strong lensing systems. In our approach, we use \texttt{Lenstronomy} to derive the magnification model from solving the lens equation with the lensed image positions, while allowing the image fluxes in the reconstruction to vary freely, thus the flux ratios between images are not constrained to match the predicted magnification model, which allows the fitted model to accommodate flux ratio anomalies. Yet tracing the physical origins for such flux ratio anomalies in strong lenses is a challenging task, as warned by many (e.g. \citealt{2003astro.ph..4480S}). We thus refrain from putting forth extensive investigation on this issue. The best-fit values of important model parameters are summarized in Table~\ref{tab:para_combined}. The results suggest an Einstein radius of 0.44$''$.

\begin{figure*}
\label{fig.jwst2_ml}
    \centering
    \includegraphics[width=1.005\linewidth]{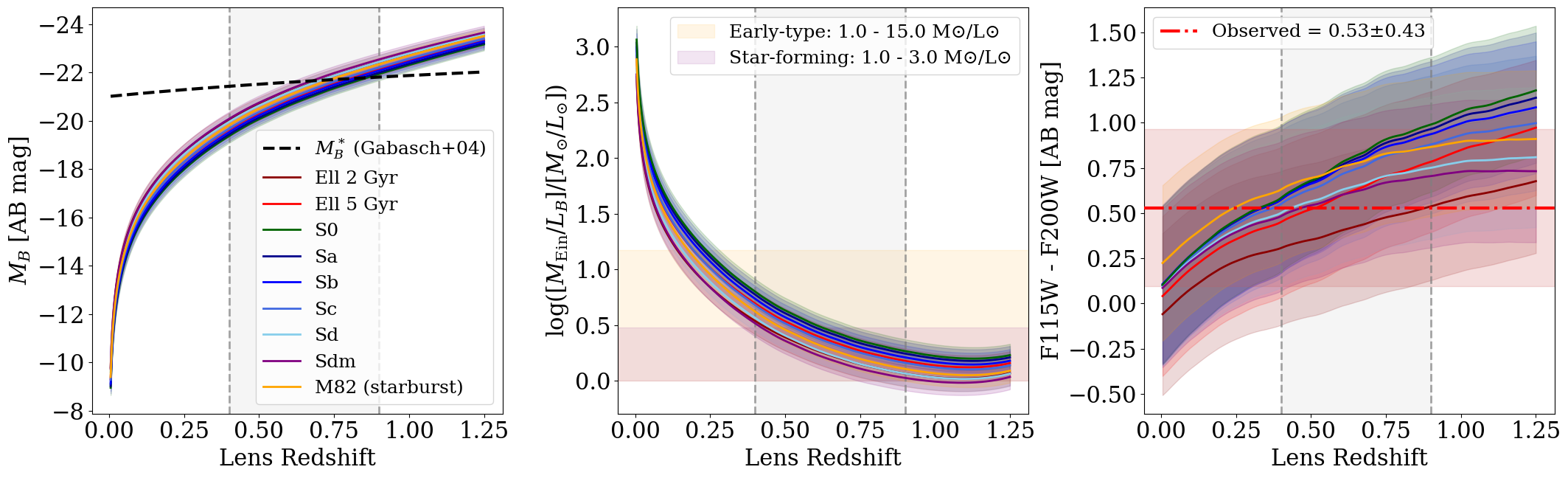}
    \caption{Photometry analysis of J0813+2545's lens. Panels are arranged in the same convention as in Fig.~\ref{fig.jwst1_ml}. Range of most probable redshift for this lens is between $0.4 < z < 0.9$.}
\end{figure*}

\subsubsection{J0803+3908}
Fig.~\ref{fig.jwst1_b} shows the NIRCam images of J0803+3908 in the F115W, F200W, F277W, and F356W filters. It includes five point sources, including four bright lensed images and a dim foreground lens that exhibits a circular shape.
A faint Einstein ring exists as well, which is more apparent in the long-wavelength filters than in the short ones. Two lensed images on the northwest of the images sit closely to each other with the one on the east being the brightest. The dimmest one is at the lower left of the field.

Fig.~\ref{fig.jwst1_f} lays out the \texttt{Lenstronomy} fit for J0803+3908 in the F200W band. We choose this band to be the most informative for i) the short-wavelength filter has a better angular resolution than the long-wavelength filters, and ii) the lens flux is larger than that in F115W. The reconstructed image is visually similar to the original observation, with a best-fit $\chi^2_{\nu} = 3.87$. There are significant ($\ge 6\sigma$) residuals around each of the four lensed images, with a concentric ring-like pattern. This implies that the PSF model we extract from the field star might be suboptimal to fully characterize the PSF structures of the observed images.
The sample field star from which we extract the PSF model is dimmer than all four lensed images, potentially causing the model to have insufficient information on diffraction spikes and speckles to model the brighter images. The two top images being too close to each other might have led to partial overlap of their PSFs, which could make the fitting even more challenging. We have experimented with other options, such as empirical PSF models from other works or simulated PSF models generated by STPSF \citep{2012SPIE.8442E..3DP}, which all struggle to give a better portrait of the complicated diffraction pattern of NIRCam. Nonetheless, the high residuals are near the point sources, and the rest of the fit is reliable. 
The fitted magnification model, similar to that for J2218-3322, shows a noticeable discrepancy with the observed flux ratios between images and might be not reliable. The other best-fit parameters are in Table~\ref{tab:para_combined}. The fit yields a Einstein radius of 0.58$''$.

\subsubsection{J0813+2545}
Fig.~\ref{fig.jwst2_b} illustrates the NIRCam images of J0813+2545 in the four NIRCam filters, displaying a dim central lensing object surrounded by four brighter lensed images. The lens of this system is visually distinguishable in all four filters, enclosed by a dim Einstein ring. 
Two overlapping bright images are located on the southwest of the image field.
In the two long-wavelength filters, pixels at the centers of quasar image are very bright and saturated, causing them to be blanked out by the data reduction pipeline and are non-recoverable.

Fig.~\ref{fig.jwst2_f} demonstrates the numerical fit results for J0813+2545 from \texttt{Lenstronomy} in the most informative F200W band. The image reconstruction yields a best-fit $\chi^2_{\nu} = 4.56$. We observe the ring-like patterns near point sources,
indicative of a similar PSF mismatch.
The positive residual patterns spanning a wide region on the top of the field is indicative of potential issue of excessive background noises. 
Fitted values of model parameter for J0813+2545 are in Table~\ref{tab:para_combined}.
This system has a best-fit Einstein radius of 0.49$''$.

\subsection{Constraints on Lens Redshift} 
As no spectroscopic measurements on the lensing galaxies are available at present, we utilize our \texttt{Lenstronomy} fit results to constrain the redshift range for each of the lenses by examining properties of the lensing galaxies, namely mass, luminosity, and color, as a function of redshift, to determine the redshift at which these quantities have the most probable values.

The total mass (for both baryonic and dark matter) enclosed within the Einstein radius $\theta_E$ of a strong lens $M_{\rm Ein}$ is:
\begin{equation}
\label{eq3}
    M_{\rm Ein} = \frac{c^2}{4G}\frac{D_L D_S}{D_{LS}}\theta_E^2,
\end{equation}
where $D_L$, $D_S$, $D_{LS}$ are the angular diameter distance to the lensing galaxy, the source quasar, and that between the two, respectively, all of which are dependent on lens redshift $z_{\rm L}$ \citep{2024SSRv..220...12S}. Therefore, we can evaluate the total enclosed mass as a function of redshift $M_{\rm Ein}(z_{\rm L})$.

In addition to the enclosed mass, given the light profile fitted by \texttt{Lenstronomy}, we can evaluate the luminosity enclosed within a lensing galaxy's Einstein radius as a function of $z_L$. For each lens, we reconstruct its surface brightness profile with the best-fit light model parameters. Then, 
we integrate the surface brightness within the Einstein ring to obtain the total enclosed spectral flux density in unit of Jansky. For the HST target, as it was observed in the F475W and F814W filter, we obtain the enclosed flux densities in both bands. To derive the corresponding total luminosity, we need to integrate the flux density over the lensing galaxy's spectral energy distribution (SED). Since the exact Hubble type of the lenses are unknown, we select 10 candidate SED templates of different galaxy types from the SWIRE template library \citep{1998ApJ...509..103S}, including two elliptical galaxies with a synthetic stellar population at 2 and 5 Gyr, six spiral galaxies of different types (Sa, Sb, Sc, Sd, S0, Sdm), as well as one starburst galaxy M82, all assuming a \cite{1955ApJ...121..161S} initial mass function. For all SEDs, we calculate synthetic flux densities in two filters and normalize to the observed fluxes in these two filters using the least squares method.
We integrate the SEDs over the rest-frame B band (4000 - 5000 $\rm \AA$), a common benchmark for galaxy studies to obtain the enclosed flux of a lens, and subsequently derive the enclosed luminosity $L(z_{\rm L})$ as a function of lens redshift.

\begin{figure*}
\label{fig.sl}
    \centering
    \includegraphics[width=1.0\linewidth]{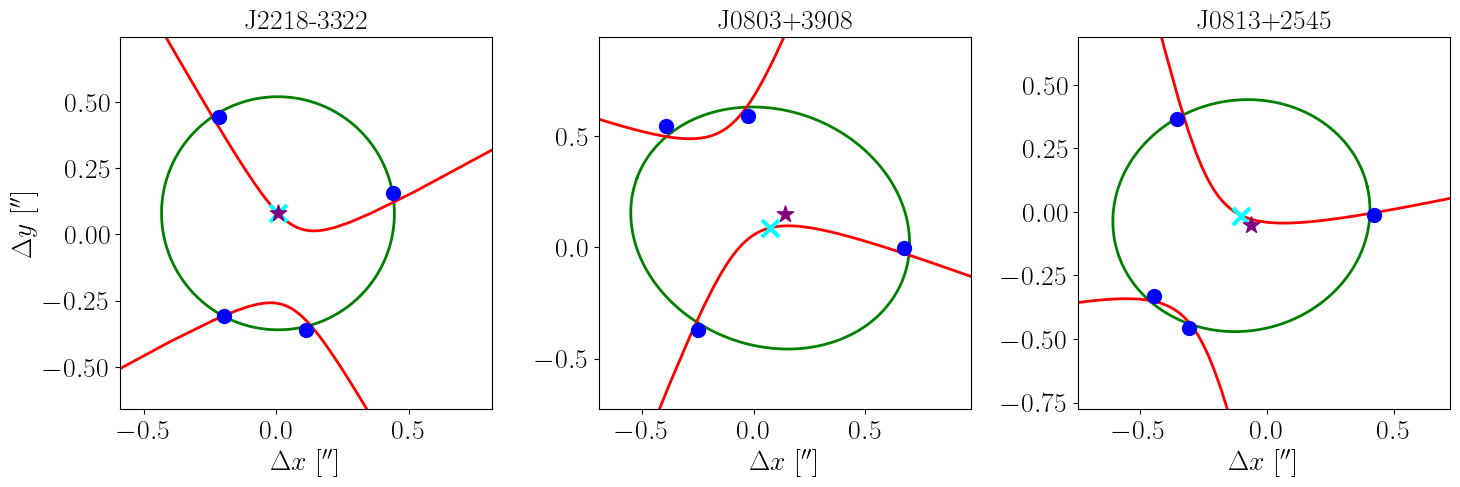}
    \caption{Fitting result for the three lensing systems in this work with the alternative model. Blue dots represent the lensed quasar images. Cyan crosses represent the fitted source position. Purple stars represent the fitted position of the center of the lensing potential.} 
\end{figure*} 

Now for every lens, we have both $M_{\rm Ein}(z_{\rm L})$ and $L(z_{\rm L})$. Thus we can evaluate the mass-to-light ratio $[M_{\rm Ein}/L](z_{\rm L})$ of a lens enclosed within the Einstein radius as a function of $z_{\rm L}$. 
In addition, we have the synthetic colors of templates as a function of $z_L$. 
We compare it to the observed color of the lens and examine at what redshift(s) would the model-dependent color match the observed value.

We also evaluate the potential systematic uncertainties associated with the photometry results. To obtain a model-independent sanity check for lens galaxy colors, we develop an independent pipeline that fits the lens light with phenomenological models (e.g. 2D Lorentzian profile, comparable to a PSF-convolved Sérsic; this also accounts for the potential systematics induced by the PSF mismatch). We intend to be as agnostic as possible about the structural parameters of the lens and just double-check whether the flux given by \texttt{Lenstronomy} is physical. One potential pitfall of \texttt{Lenstronomy} fits is the software does not include a background in the lens light model, i.e. it assumes that the observed flux is entirely due to the source and the lens. While realistically those pixels might be subject to various contamination, including noise fluctuations, bright PSF spikes or source galaxy light leaking into the Einstein ring, etc. The flux/luminosity derived by \texttt{Lenstronomy} thus could have overestimated the lens light. Our custom pipelines examine the potential systematics induced by this assumption by fitting models both with and without a background. The former consists of a uniform background accounting for contaminating fluxes and a central profile for lens light. Derived flux from the central profile marks the lower bound of total enclosed flux within Einstein radius. The without-background model, inheriting similar assumptions with \texttt{Lenstronomy}, sets the upper bound. We set the best estimate and associated uncertainty of lens flux to be the mean and standard error of these measurements at the 95\% confidence level.

Fig.~\ref{fig.hst_ml}, ~\ref{fig.jwst1_ml}, and ~\ref{fig.jwst2_ml} present the photometry results for the lens of J2218-3322, J0803+3908, and J0813+2545, respectively. The dotted black line in each B-band lens luminosity $M_B(z_{\rm L})$ plot displays an empirical relation between the B-band characteristic luminosity $M^*_B$ and redshift $z$ as $M^{*}_{B}(z) = -20.92-1.25 \rm \ ln(1 + z)$ by \cite{2004A&A...421...41G}, as a reference for how is the lens luminosity compared to the “representative” value at a given redshift. We stop plotting out to higher redshifts once a lens' luminosity surpasses local $M_B^*$ value by $\sim 2 - 3$ magnitudes. The colored boxes in the enclosed mass-to-light ratio $[M_{\rm Ein}/L_B](z_{\rm L})$ plots mark a range of typical mass-to-light ratios for different galaxy types cited from the literature (e.g. \citealt{2002MNRAS.334..721G,2007ApJ...668..756V,2013MNRAS.432.1709C}). Those references report the dynamical mass-to-light ratios within 1 $R_e$ of galaxies, that is, a few kpc within a galaxy's center. The Einstein radius of a lens corresponds to different physical radii at different redshifts. For the lenses in this work, within the possible redshift range their Einstein radii are in between $\sim 1 - 3.5$ kpc, which are compatible with the reference values. We choose the bounds on $M/L_B$ to be loose. With Equation 23 in \citet{2016ARA&A..54..597C} and typical effective velocity dispersion $\sigma_{\rm eff}$ $\sim$ 50 - 400 km/s for early-types (e.g. \citealt{2008ApJ...684..248B, 2012MNRAS.419..656C, 2016ARA&A..54..597C}), we estimate a plausible $M/L_B$ range of $\sim$ 2 - 14 solar units. Indeed, galaxies with high $M/L_B$ values near or even above 15 solar units are intrinsically rare; they are the most massive cluster-scale systems, while the majority of observed galaxies sits well below this bound \citep{2020ARA&A..58..577S}. We want to be as conservative about the lens redshifts as possible, since strong lensing galaxies might well be different from the field galaxy population, and especially for these small separation lenses in our sample which are yet poorly understood. Thus we push the bounds to slightly above the limit available in observations. We use $[M_{\rm Ein}/L_B](z_{\rm L})$ and the redshift-dependent color ($\rm [F475W - F814W]$$(z_{\rm L})$ for HST target and $\rm [F115W - F200W]$$(z_{\rm L})$ for JWST targets) to constrain the most possible range of redshift for a lens: if i) the mass-to-light ratio of a certain galaxy model is within a reasonable range at a certain redshift, and ii) the color of the model is compatible with the observed value (model \& observed color directly meet, or error bars overlap) at that redshift, we then conclude the lens could be that type of galaxy at that redshift. We report the constrained redshift range for each lens to be: J2218-3322: $0.5 < z < 1.2$; J0803+3908: $1.0 < z < 1.5$; J0813+2545: $0.4 < z < 0.9$. Implications from these constraints are discussed in Sec.~\ref{sec:dis}.

\begin{table*}
\centering
\label{tab:flux_ratios}
\begin{tabular}{llccc}
\hline
System & Ratio & \texttt{Lenstronomy} light & \texttt{Lenstronomy} mass & Witt-Wynne \\
\hline
\multirow{4}{*}{J2218-3322}
& A/B & $0.855^{+0.021}_{-0.026}$ & $0.512^{+0.030}_{-0.028}$ & $1.034^{+0.021}_{-0.033}$ \\
& B/B & $1.000$ & $1.000$ & $1.000$ \\
& C/B & $0.222^{+0.017}_{-0.019}$ & $0.444^{+0.021}_{-0.021}$ & $1.118^{+0.040}_{-0.056}$ \\
& D/B & $0.559^{+0.023}_{-0.025}$ & $1.130^{+0.064}_{-0.058}$ & $1.081^{+0.054}_{-0.065}$ \\
\hline
\multirow{4}{*}{J0803+3908}
& A/D & $0.291^{+0.004}_{-0.009}$ & $0.235^{+0.048}_{-0.041}$ & $4.873^{+6.097}_{-2.228}$ \\
& B/D & $0.547^{+0.002}_{-0.003}$ & $0.466^{+0.029}_{-0.023}$ & $1.336^{+0.085}_{-0.059}$ \\
& C/D & $0.732^{+0.001}_{-0.002}$ & $1.010^{+0.081}_{-0.040}$ & $5.339^{+8.124}_{-1.498}$ \\
& D/D & $1.000$ & $1.000$ & $1.000$ \\
\hline
\multirow{4}{*}{J0813+2545}
& A/D & $0.444^{+0.011}_{-0.021}$ & $1.519^{+0.739}_{-0.499}$ & $1.189^{+1.056}_{-0.594}$ \\
& B/D & $0.091^{+0.064}_{-0.033}$ & $0.206^{+0.064}_{-0.059}$ & $0.903^{+0.612}_{-0.395}$ \\
& C/D & $0.262^{+0.030}_{-0.025}$ & $0.341^{+0.065}_{-0.057}$ & $0.658^{+0.361}_{-0.376}$ \\
& D/D & $1.000$ & $1.000$ & $1.000$ \\
\hline
\end{tabular}
\caption{Flux ratios of the four lensed quasar images in each lensing system, derived from \texttt{Lenstronomy} light model fit, \texttt{Lenstronomy} mass model fit, and the Witt-Wynne model fit. For each system the flux ratios are normalized to the brightest image.}
\end{table*}

\subsection{Alternative Lensing Model}
\label{subsec:alt}
In this section we explore an alternative model for our systems -- specifically, the singular isothermal elliptical lensing potential, one of the simplest models that produces a quadruple lens, following prescriptions by \citet{witt96, wynn18, luht21, sche24}. In this model, the two-dimensional lensing potential is 
\begin{equation}
\psi(\theta_1,\theta_2)=b\sqrt{\theta_1^2+\theta_2^2/q^2}
\end{equation}
using the same notation as equation (\ref{eq:convergence}), i.e., $\theta_1$ and $\theta_2$ are measured along the major and minor axes of the lensing potential from the center of the lens. 
The convergence for this model $\kappa(\theta_1,\theta_2)\propto (\theta_1^2+\theta_2^2)/(\theta_1^2+\theta_2^2/q^2)^{3/2}$, i.e., similar to but not exactly the same as the convergence of the power-law ellipsoid with $\gamma=2$ we explored with \texttt{Lenstronomy} (for which the analytical lensing potential is given by \citealt{schn06}). 

For this potential parameterized by the overall normalization $b$ and ellipticity $q$, 
the four images must lie on an ellipse (``Wynne's ellipse", \citealt{wynn18}) centered on the projected (original, unlensed) source position $x_S, y_S$. They must also lie on a rectangular hyperbola, i.e., a hyperbola whose asymptotes are orthogonal to each other, which passes through both the center of the lens $x_L, y_L$ and the projected source position $x_S, y_S$ (``Witt's hyperbola", \citealt{witt96}). Both the hyperbola and the ellipse have the same position angle $\theta_L$ which is the position angle of the long axis of the potential distribution, and they both have $q$ as a shape parameter.
Additionally, the ellipse depends on the normalization of the lensing potential $b$. Therefore, the four positions of the images provide 8 data points ($x_i,y_i$, $i=1...4$) to construct a 7-parameter model of the best-fitting Witt's hyperbola and Wynne's ellipse which yields all the parameters of the lensing potential and the unlensed position of the source. 

In our implementation of the isothermal elliptical potential fit, we take the best-fitting image positions from our \texttt{Lenstronomy} fits and minimize the sum of the squares of the distances from the images to Wynne's ellipse and Witt's hyperbola to obtain the best-fitting model parameters. We require the Witt's hyperbola to pass through the unlensed source position, but not necessarily the center of the lensing potential. This is inspired by the procedure by \citet{sche24} who indicate that the center of lensing potential could be displaced from the hyperbola due to physical reasons that are discussed in Sec.~\ref{subsec:models}. We experiment with two other fitting approaches, either requiring that the centroids of both the potential and the unlensed source position are on the hyperbola, or requiring none of the centroids to be on the hyperbola. These three recipes in principle should all reproduce the same lensing geometry, conditioned on that the true lensing potential is similar to the assumed analytical model. We find that only the first approach (fixing centroids of the source but not potential on the hyperbola) yields reasonable fits. This could imply that the lensing potential might be not well described by the model.

Overall we find that the positions of the images are fit well for all three targets in this work as shown in Fig.~\ref{fig.sl}, reproducing them within $\sim 0.01 - 0.03''$ from the Wynne's ellipse or the Witt's hyperbola, which are $\sim 3\% - 7\%$ of their lenses' Einstein radii. The structural parameters yielded from this model are not consistent with the values fitted by \texttt{Lenstronomy}. The lens and source positions fitted by each model differ by $\sim 0.05''$. The axis ratios from the \texttt{Lenstronomy} light model differ with those from the Witt-Wynne model by at most $\sim 0.3$, and the position angles differ by at most $\sim 60^{\circ}$. In contrast, the differences between axis ratios of the mass model in the Lenstronomy and Witt-Wynne models are at most $\sim 0.05$, and that between the position angles are at most $\sim 20^{\circ}$.

Such discrepancies between the two mass models could be another implication that the true lensing potential is more complicated than that assumed by the analytical model. The potential adopted by \texttt{Lenstronomy}, which reproduces better fits of lensed image positions, could be more loyal to reality than the Witt-Wynne model. Besides, the Witt-Wynne model is essentially geometric: it fits the structural parameters of the lensing potential using only the lensed image positions. \texttt{Lenstronomy} fits more information (mass \& light, quasar host galaxy etc.) simultaneously with the input image. While the lensing potentials adopted by both models are similar, the data being fitted might not be necessarily the same. Moreover, the structural parameters of the lenses may not be well constrained. As stated in Sec.~\ref{subsec:rob}, parameters in \texttt{Lenstronomy} fits show a ``hierarchy of errors" with the ellipticity terms (axis ratio and position angle) being the most poorly determined. 

Table.~\ref{tab:flux_ratios} presents the flux ratios on the three quads derived from the \texttt{Lenstronomy} light model, \texttt{Lenstronomy} mass model, and the Witt-Wynne model, respectively. These comparisons reveal two intriguing observations: i) \texttt{Lenstronomy} mass and light model give flux ratios that on average differ by a factor of $\sim 1.9$ for all three quads, whereas that between \texttt{Lenstronomy} mass and Witt-Wynne on average differ by a factor of $\sim 4.7$; ii) \texttt{Lenstronomy} models and Witt-Wynne give very different relative ratios for certain images, e.g. for J0803+3908, \texttt{Lenstronomy} indicates A to be dimmer than B, while Witt-Wynne suggests the opposite. The average difference between flux ratios yielded by the \texttt{Lenstronomy} models are at a level consistent with those found in classical anomalous systems such as B1422+231, where smooth models fail to reproduce the observed ratios by factors of $\sim 1.5–2$ (\citealt{1998MNRAS.295..587M, 2002A&A...388..373B}). The average difference between the \texttt{Lenstronomy} mass and the Witt-Wynne model, on the other hand, is far beyond that range. This serves as an indicative metric for quantifying the systematic uncertainties in our analysis due to assumptions of a particular functional form for the lensing potential.

\section{Discussion} \label{sec:dis}
\subsection{Classifications \& Properties of Lenses}
Strong lensing galaxies detected until now are biased towards a specific type: bright, compact, massive early-type galaxies at an intermediate redshift ($z \leq 1$). A galaxy can act as a strong lens only when its surface mass density reaches or exceeds the critical threshold $\sum_{\rm crit}$. This condition favors galaxies with more mass and more compact geometry. Besides, major brightness-limited lens surveys tend to select more luminous galaxies \citep{2023A&A...678A...4S, 2024A&A...690A.325S}. Due to these selection effects, compact massive early-type galaxies overwhelmingly dominate the documented population of strong lensing galaxies \citep{2009ApJ...705.1099A, 2010A&A...517A..25S}. Late-type/star-forming galaxies, while more numerous, are far less common among lenses, accounting for only $\sim 10\% - 20\%$ of all lenses with a substantially lower lensing cross-section than early-types \citep{1998ApJ...495..157K, 2015ApJ...811...20C, 2023A&A...678A...4S}. Out of 40 lensing galaxies in quadruply lensed quasars with known lens redshift documented as of October 2025 (see Appendix), 33 have classifications of galaxy types reported, among which 27 have been classified or suspected as early-type galaxies, while only 6 are claimed to be late-type/spiral galaxies.

The three lenses in our sample are likely early-type galaxies as well. We measured the $\rm S\acute{e}rsic$ index for each lens to be 4.24, 4.10, and 3.37, i.e. all consistent with a de Vaucouleurs $n_{\rm S\acute{e}rsic} = 4$ light profile for elliptical/early-types. 
The best-fit effective radii for these lenses are 0.30$''$, 0.35$''$, and 0.45$''$. At their respective redshift ranges, the corresponding physical distances are around 1.5 - 3.5 kpc. For massive early-types of $\sim 10^{10} - 10^{11}\ \rm M_{\odot}$ at $z \leq 1$, the typical $R_e$ is in between 1 - 5 kpc (e.g. \citealt{2013MNRAS.432.1862C,2014ApJ...788...28V}). The three lenses thus fit comfortably into the mass-size plane for early-type galaxies. The stellar velocity dispersions inferred from the SIE fits for the lenses range across $\sim 150 - 250$ km/s. At lower redshifts ($z \leq 1$) the bulk of early-type galaxies exhibits velocity dispersions in between $\sim 50 - 400$ km/s (e.g. \citealt{2008ApJ...684..248B, 2012MNRAS.419..656C, 2016ARA&A..54..597C}). The lenses in this work lay on the middle to lower end of that range, corresponding to $L^*$ and sub-$L^*$ early-type galaxies, consistent with the lens luminosities out of the photometric analysis. This also implies that the lenses are more typical of field ellipticals or small-group brightest galaxies rather than massive cluster-scale lenses with high velocity dispersions $\geq$ 300 km/s.

\begin{figure*}
\label{fig.theta_z}
    \centering
    \includegraphics[width=1.0\linewidth]{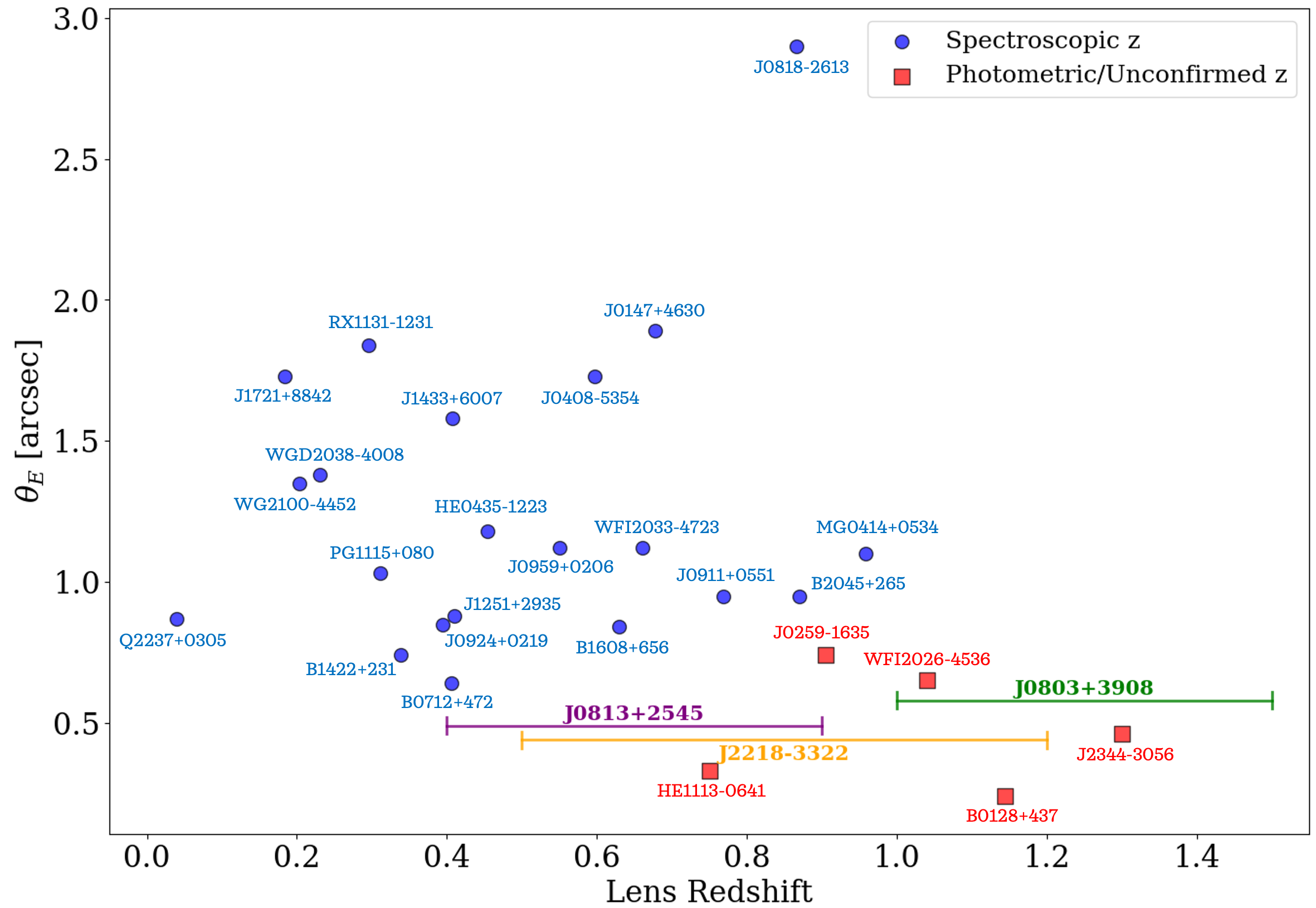}
    \caption{Lens redshifts and Einstein radii $\theta_E$ of lensing galaxies of known quadruply lensed quasars. Previously discovered objects are plotted with dots and squares. The three targets in this work are plotted with barred colored horizontal lines. See Appendix for a complete list of objects shown and corresponding reference.} 
\end{figure*}

J2218-3322 exhibits a significant redness. While dust extinction and/or external reddening cannot be ruled out, it is a strong indicator of a passively evolving galaxy i.e. early-type, consistent with the previous classification. J0803+3908 and J0813+2545 show bluer colors, which is less common for early-type galaxies that have evolved and show little to no ongoing star formation. Certain scenarios might account for such an observation. Some elliptical and S0s contain substantial cold gas reservoirs, either from recent gas-rich mergers or ongoing accretion, that boost star formation rates (e.g. \citealt{2007MNRAS.377.1795C}). Recent major mergers can trigger starbursts in early-types, particularly in the transition phase, causing a temporarily blue-ward shift in color (e.g. \citealt{2017A&A...598A..45G}). 

\subsection{Quasar Host Galaxies}
We are conservative about making statements regarding host galaxies for the lensed quasars, especially for the JWST targets due to the PSF mismatch. That said, there is an interesting pattern for both JWST objects in Fig.~\ref{fig.jwst1_f} and Fig.~\ref{fig.jwst2_f} that the residuals along the Einstein ring show reverse signs at different segments. For example, in Fig.~\ref{fig.jwst1_f}, the residuals between lensed image B and D at the northeast of the figure have a positive sign, while the residuals between image A and C as well as A and B are largely negative. This spatial variation of residual sign could imply that the source quasar is not located at the center of the host galaxy, because the \texttt{Lenstronomy} fits assume that the quasar and the host are co-centered.

\subsection{Degeneracy of Lensing Models}
\label{subsec:models}
The comparison between results given by \texttt{Lenstronomy} and the alternative analytical model by \cite{luht21} highlights the variety/degeneracy in lensing models. It is long known that the observed configuration of a strong lensing event i.e. lensed image positions, flux ratios etc. could be reproduced by different lensing models (e.g. the mass-sheet degeneracy; \citealt{1985ApJ...289L...1F, 2013A&A...559A..37S}). Although both \texttt{Lenstronomy} and the analytical model adopt an isothermal lensing mass/potential (with external shear), their underlying assumptions are not identical. One key difference is that the analytical model does not require the lensing potential to be centered on the lensing galaxy unlike what most lens modeling codes like \texttt{Lenstronomy} do \citep{sche24}. Rather, the model predicts that the lensing potential center lies on the Witt's hyperbola somewhere with a small but non-zero offset (typically $\sim 0.01''$) to the lensing galaxy. This implies either: i) the potential is intrinsically displaced from the lensing galaxy (as seen in many bright cluster galaxies e.g. \citealt{2011MNRAS.410..417S}), or ii) the true potential is more intricate than the model's prediction. We observe such offsets in the fits for all three targets in this work. The discrepancy between structural parameters by this model and the \texttt{Lenstronomy}-fitted results seems to favor the first scenario of a difference in the geometry of lensing potential and observed shape for each lens.

The striking discrepancies in flux ratios by the models further highlights the key challenges in selecting the right model to model the fluxes of strong lenses. It seems to imply that even though the potentials adopted by the two models are similar, a small difference in the potentials could lead to considerable deviations in fluxes. On top of that, microlensing could also have effects: \cite{2021ApJ...922...70W} report an rms microlensing effect of $\sim$ 0.5 mag to lensed quasar images with a 95\% confidence range at least twice as large, based on statistical modeling of quasar microlensing across a range of lens configurations, complicating the flux ratio evaluations even more. All these insights emphasize the complexity of realistic galaxies.

\subsection{Comparison with Known Quad Lenses}
Fig.~\ref{fig.theta_z} illustrates the distribution of lens redshift and Einstein radius of lensing galaxies for quadruply lensed quasars and demonstrates how our lenses fit into the population. It contains 26 of all 27 lensing galaxies in known quads with published measurements on lens redshift by far (we exclude J1004+4112 which is a cluster lens and thus has a large Einstein radius of $8.14''$) along with the three lenses in this work. The lens redshifts are scattered across $z = 0.0 \sim 1.4$ with a mean at $z = 0.592$. The Einstein radii $\theta_E$ vary between $0.25'' \sim 3.00''$ with a mean of $1.12''$, while the majority are above $0.60''$. Our three lenses are located at the lower end of the distribution of $\theta_E$. 
In terms of redshift, the expected redshift ranges for both J2218-3322 and J0813+2545 lie within this distribution well. J0803+3908 is on the higher end of lens redshift in this quad lens sample set along with very few others at $z \ge 1$.

It is noteworthy that lenses with similar Einstein radii with our samples in  Fig.~\ref{fig.theta_z} tend to have photometric or uncertain redshifts. This highlights the observational challenge for obtaining spectroscopic measurements on small-separation quads and lenses, especially with ground-based instruments. Atmospheric seeing limits the resolvability of ground telescope on subarcsec separation sources. That said, some current space observatories such as JWST NIRSpec already offer sufficient capacity probing these targets. Upcoming space missions such as Euclid and Roman \citep{2020RNAAS...4..190W, 2025A&A...696A.214P} would probe deeper into this largely unexplored regime, resolving lenses with smaller separations (i.e. lower enclosed mass) and/or at higher redshift. Such lenses at earlier cosmic epochs might not necessarily be similar to the contemporary population. As we approach cosmic noon, field galaxies contain more young, bright galaxies with intense star formation. While massive early-types would still be the dominating type due to their more favorable surface density configurations, we can expect to detect more late-type/edge-on and lower-mass galaxies as strong lenses. These objects would greatly advance our understanding of mass assembly history of early galaxies.

\section{Conclusion} \label{sec:con}
We present optical imaging for three quadruply lensed quasars at cosmic noon: J2218-3322 with HST/WFC3, MQJ0803+3908 with JWST/NIRCam, and SDSSJ0813+2545 with JWST/NIRCam. With numerical modelings of these systems using \texttt{Lenstronomy}, we retrieve vital information about each of these quads, and perform aperture photometry to constrain the possible redshifts to each lensing galaxy. Our key results include:
\begin{enumerate}
    \item \texttt{Lenstronomy} yields compatible image reconstructions for all three systems, with best-fit $\chi^2_{\nu} = 1.21, 3.87,$ and $4.56$. The sub-optimal fits for the two JWST quads are mostly due to mismatch of PSF models caused by the complex diffraction patterns of NIRCam.
    \item The three lenses are all compact systems with Einstein radii of 0.44$''$, 0.58$''$, and 0.49$''$. The most probable redshift range for each one is $0.5 < z < 1.2$, $1.0 < z < 1.5$, and $0.4 < z < 0.9$, respectively.
    \item The three lenses are all likely early-type galaxies given their $n_{\rm S\acute{e}rsic} \sim 4$ light profile and modest $R_e$ around a few kpc. Compared to other lensing galaxies for quadruply lensed quasars, they have an uncommonly compact Einstein radius below 1$''$. These objects vary drastically in terms of mass, color, and luminosity, together forming a diverse subset of the lensing galaxy population.  
\end{enumerate}
Our work exhibits a rich ecosystem of single-galaxy lenses where galaxies with distinct characteristics are all capable of producing strong gravitational lensing, and peeks into the exciting realm of subarcsec separation strong lenses that remains under-studied so far.
Upcoming surveys on next-generation instruments, including both ground-based observatories such as the Legacy Survey of Space and Time program on the Vera C. Rubin Observatory \citep{2025RSPTA.38340117S} and space telescopes like the Euclid and Roman missions could expectedly expand the sample size of strong lenses by orders of magnitude, while each of them would be observed with unprecedented resolutions and depths \citep{2010MNRAS.405.2579O}. Our work showcases a workflow on dealing with the rich complex details from each strong lensing system, which can yield new insights that will lead to extraordinary advancements from the current understandings on single-galaxy strong lensing, as well as the broader paradigm of galaxy formation and evolution.

\section*{Acknowledgments}
Observations used for this work were conducted with the NASA/ESA \textit{Hubble Space Telescope} and \textit{James Webb Space Telescope} and acquired from the Space Telescope Science Institute, operated by the Association of Universities for Research in Astronomy, Inc., under NASA contract NAS 5-26555. Support for program JWST-GO-04204 was provided by NASA through a grant from the Space Telescope Science Institute, which is operated by the Association of Universities for Research in Astronomy, Inc., under NASA contract NAS 5-03127. The authors thank the anonymous referee for insightful comments. M.L. acknowledges support from Space@Hopkins. N.L.Z. thanks the Institute for Advanced Study in Princeton, NJ for hospitality and support during regular visits, and thanks P. Schechter for many useful conversations. X.L. acknowledges support by NSF grant AST-2108162. This work utilizes the JWST Calibration Pipeline \citep{bushouse_2025_17515973}, \texttt{Lenstronomy} \citep{2018PDU....22..189B}, \texttt{emcee} \citep{2013PASP..125..306F}, \texttt{fastell} \citep{1999ascl.soft10003B}, \texttt{numpy} \citep{2020Natur.585..357H}, \texttt{scipy} \citep{2020NatMe..17..261V}, \texttt{astropy} \citep{2018AJ....156..123A}, \texttt{photutils} \citep{larry_bradley_2025_14889440}, and \texttt{matplotlib} \citep{2007CSE.....9...90H}. 

\section*{Data}
All the {\it HST} and {\it JWST} data used in this paper can be found in MAST: \dataset[10.17909/921v-5j15]{http://dx.doi.org/10.17909/921v-5j15}.

\bibliography{ref.bib}{}
\bibliographystyle{aasjournal}

\clearpage
\section*{Appendix}
\label{appendix}
\begin{table}[!ht]
List of quadruply lensed quasars with known lens redshift by October 2025 (excluding the three objects from this work):

\centering
\label{tab:quad_lens_galaxy_types}
\begin{tabular}{lcccclc}
\hline
Name & $z_{\rm lens}$ & $z_{\rm source}$ & $\theta_{\rm E}$ & Lens Type & $ $ Reference  & Reference \\
     &                &                   & ($''$)         & &(discovery) & ($\theta_{\rm E}$ and/or Type) \\
\hline
B0128+437        & 1.145$^{\rm *}$ & 3.124 & 0.24 & Spiral/S0 & \citet{2000MNRAS.319L...7P} & \citet{2010ApJ...716L.185L} \\
J0134$-$0931     & 0.765 & 2.220 & --- & Pair (S + S) & \citet{2002ApJ...564..143W} & \citet{2018ApJ...864...73W} \\
J0147+4630       & 0.678 & 2.377 & 1.89 & Early-type & \citet{2017ApJ...844...90B} & \citet{2023MNRAS.518.1260S} \\
J0214$-$2105     & 0.450$^{\rm *}$ & 3.240 & --- & --- & \citet{2018RNAAS...2...42A} & \citet{2018RNAAS...2...42A} \\
J0230$-$2130     & 0.523 & 2.162 & --- & Pair (E + S?) & \citet{Wisotzki99} & \citet{Eigenbrod06} \\
J0259$-$1635     & 0.905$^{\rm *}$ & 2.160 & 0.74 & --- & \citet{2018RNAAS...2...21S} & \citet{2023MNRAS.518.1260S} \\
J0408$-$5354     & 0.597 & 2.375 & $1.73$ & Early-type & \citet{2017MNRAS.472.4038A} & \citet{2017ApJ...838L..15L} \\
MG0414+0534      & 0.958 & 2.639 & $1.10$ & Elliptical & \citet{1992AJ....104..968H} & \citet{2009ApJ...697..610M} \\
HE0435$-$1223    & 0.454 & 1.693 & $1.18$ & Elliptical & \citet{Wisotzki02} & \citet{2017MNRAS.465.4895W} \\
B0712+472        & 0.406 & 1.339 & $0.64$ & Late-type & \citet{1998MNRAS.296..483J} & \citet{2017MNRAS.469.3713H} \\
J0818$-$2613     & 0.866 & 2.155 & 2.90 & Early-type? & \citet{Mozumdar23} & \citet{2023MNRAS.518.1260S} \\
J0911+0551       & 0.769 & 2.763 & 0.95 & Early-type? & \citet{Bade97} & \citet{2015MNRAS.454..287J} \\
J0924+0219       & 0.394 & 1.524 & $0.85$ & Elliptical & \citet{Inada03} & \citet{Inada03} \\
J0959+0206       & 0.551 & 3.140 & 0.74 & --- & \citet{Anguita2009} & \citet{2015ApJ...806..185C} \\
J1004+4112       & 0.680 & 1.734 & $8.14$ & Elliptical + cluster & \citet{Inada03} & \citet{Oguri2010} \\
FSC10214+4724    & 0.900$^{\rm *}$ & 2.286 & --- & Elliptical & \citet{Rowan-Robinson1991} & \citet{1995ApJ...450L..41B} \\
J1042+1641       & 0.599$^{\rm *}$ & 2.500 & --- & Early-type & \citet{2023ApJ...943...25G} & \citet{2023ApJ...943...25G} \\
HE1113$-$0641    & 0.750$^{\rm *}$ & 1.235 & 0.33 & --- & \citet{2008AJ....135..374B} & \citet{2008AJ....135..374B} \\
PG1115+080       & 0.311 & 1.722 & $1.03$ & Pair (E + E?) & \citet{1987ApJ...312...45C} & \citet{1998ApJ...509..551I} \\
RXJ1131$-$1231   & 0.295 & 0.658 & $1.64$ & Elliptical & \citet{Sluse2003} & \citet{Suyu2013} \\
J1138+0314       & 0.445 & 2.438 & --- & Early-type & \citet{Inada2008} & \citet{Inada2008} \\
J1251+2935       & 0.410 & 0.802 & $0.88$ & Early-type & \citet{Kayo2007} & \citet{Kayo2007} \\
2M1310$-$1714    & 0.293 & 1.975 & --- & Early-type & \citet{Lucey2018} & \citet{Lucey2018} \\
J1330+1810       & 0.373 & 1.393 & --- & Late-type & \citet{Oguri2008} & \citet{2016MNRAS.458....2R} \\
B1359+154        & 0.900$^{\rm *}$ & 3.235 & $1.10$ & Group (3 galaxies) & \citet{Myers1999} & \citet{Rusin2001} \\
HST14113+5211    & 0.465 & 2.811 & --- & Spiral & \citet{Fischer1998} & \citet{Ratnatunga99} \\
H1413+117        & 0.940$^{\rm *}$ & 2.560 & --- & --- & \citet{Magain1988} & \citet{Magain1988} \\
B1422+231        & 0.339 & 3.620 & $0.74$ & Early-type + group & \citet{Patnaik1992} & \citet{Kormann1994} \\
J1433+6007       & 0.407 & 2.740 & 1.58 & Early-type & \citet{Agnello2018} & \citet{2023MNRAS.518.1260S} \\
B1608+656        & 0.630 & 1.394 & $0.84$ & Pair (E + S) & \citet{Fassnacht1999} & \citet{Suyu2009} \\
J1721+8842       & 0.184 & 2.370 & 1.74 & Pair (E + E) & \citet{Lemon2018} & \citet{Dux2025} \\
B1933+503        & 0.755 & 2.638 & --- & Early-type & \citet{Sykes1998} & \citet{Sykes1998} \\
B1938+666        & 0.881 & 2.059 & --- & Early-type & \citet{Patnaik1992} & \citet{Lagattuta12} \\
WFI2026$-$4536   & 1.040$^{\rm *}$ & 2.230 &  0.65 & --- & \citet{Morgan2004} & \citet{Sluse2012} \\
WFI2033$-$4723   & 0.661 & 1.662 & $1.12$ & Elliptical + group & \citet{Morgan2004} & \citet{Sluse2012} \\
WGD2038$-$4008   & 0.230 & 0.777 & 1.38 & Early-type & \citet{Agnello2018b} & \citet{Buckley-Geer2020} \\
B2045+265        & 0.870 & 1.280 & 0.95 & Elliptical + satellite & \citet{Fassnacht1999b} & \citet{McKean07} \\
WG2100$-$4452    & 0.203 & 0.920 & 1.35 & --- & \citet{Agnello2019} & \citet{Ertl2023} \\
Q2237+0305        & 0.039 & 1.695 & $0.87$ & Spiral & \citet{Yee1988} & \citet{Wambsganss1994} \\
J2344$-$3056     & 1.300$^{\rm *}$ & 1.298 & 0.50 & Elliptical? & \citet{Schechter2017} & \citet{He2025} \\
\hline
\multicolumn{7}{l}{$^{\rm *}$ Unconfirmed or photometric redshift.} \\
\multicolumn{7}{l}{Notes: S = Spiral; E = Elliptical; S0 = Lenticular; ``Early-type" includes E \& S0 galaxies.} \\
\multicolumn{7}{l}{``---'' indicates no information available in the literature. "?" indicates the classification is not definitive.} \\
\hline
\end{tabular}
\end{table}



\end{document}